\documentclass[prb,twocolumn,showpacs,preprintnumbers,amsmath,amssymb]{revtex4}
\usepackage{graphicx}
\usepackage{dcolumn}
\usepackage{bm}
\usepackage{mathrsfs}
\usepackage{subfigure}
\usepackage{natbib}
\usepackage{color}

\begin{document}
\title{Ginzburg-Landau theory for multiband superconductors: microscopic derivation}

\author{N. V. Orlova, A. A. Shanenko, M. V. Milo\v{s}evi\'{c}, and F. M. Peeters}
\affiliation{Departement Fysica, Universiteit Antwerpen,Groenenborgerlaan 171, B-2020 Antwerpen, Belgium.}
\author{A. Vagov and V. M. Axt}
\affiliation{Institut f\"{u}r Theoretische Physik III, Bayreuth Universit\"{a}t, Bayreuth 95440, Germany.}

\begin{abstract}
A procedure to derive the Ginzburg-Landau (GL) theory from the multiband BCS Hamiltonian is developed in a general case with an arbitrary number of bands and arbitrary interaction matrix. It combines the standard Gor'kov truncation and a subsequent reconstruction in order to match accuracies of the obtained terms. This reconstruction recovers the phenomenological GL theory as obtained from the Landau model of phase transitions but offers explicit microscopic expressions for the relevant parameters. Detailed calculations are presented for a three-band system treated as a prototype multiband superconductor. It is demonstrated that the symmetry in the coupling matrix may lead to the chiral ground state with the phase frustration, typical for systems with broken time-reversal symmetry.
\end{abstract}
\pacs{74.20.De, 74.20.Fg}

\maketitle
\section{Introduction}
\label{sec:Int}

Studies of multiband (or multigap) superconducting systems, where more then one carrier band contributes to the formation of the condensate, have now more then a half-century history.\cite{suhl,mos,geilik,kresin} In the last decade clear experimental evidences of multigap condensates were observed in a rich variety of materials such as magnesium diboride,\cite{can} oxypnictides,\cite{kamihara} iron arsenides\cite{rot} and iron pnictides.\cite{paglion} The string of discoveries continues today so that the number of multiband superconductors increases almost yearly.

There are different physical mechanisms responsible for the formation of multiple carrier bands. For example, in bulk specimens the multi-gap structure can be related to the appearance of separate pockets in the Fermi surface centered around some points of the Brillouin zone. However, it was recently shown that multiband superconductivity can also arise in nanoscale specimens (e.g. in nanofilms) made of ordinary single-band superconducting materials, where the geometrical size quantization creates distinct carrier subbands.~\cite{nanofilms} These and similar findings broadened the interest in the physics of multiband coherent phenomena, and that interest has given a strong impetus to theoretical investigations. One of the  focal points of such investigations is revisiting long established and widely used theoretical models and methods in superconductivity in the context of multiband superconductors. One of such methods is the Ginzburg-Landau (GL) theory,\cite{lan} which is commonly regarded as one of the most general and yet simple approaches for conventional single-band superconductors. Surprisingly, the generalization of the GL theory to the multiband case is still a highly debated issue.

On the microscopic level a multiband superconductor is modeled by the multiband generalization of the BCS theory.\cite{suhl,mos} The corresponding GL equations are derived using a straightforward application of the original single-band recipe by Gor'kov,\cite{gork} see, e.g., Refs.~\onlinecite{gur,zhit}. In this approach superconducting gap functions in each carrier band, hereafter referred to as band gaps, are regarded as the order parameters. Similarly to the single-band case, the anomalous Green's function of each band is expanded in powers of the corresponding band gap and its spatial gradients,\cite{gork} and then the expansion is truncated to keep the same terms as in the single-band GL theory. This procedure yields a system of nonlinear GL-like equations, one for each band gap, coupled via the linear Josephson-like terms, and the corresponding multi-component functional. This is often referred to as the {\it multi-component} GL model and is widely used in the analysis of multiband superconductors.
\cite{gur,zhit,gol,babaev,babaev1,babaev2,tes,frust,carl,geurts,tanaka}

Although this formulation of the GL theory appears intuitively justified, partially by a familiar structure of the obtained equations, it possesses several fundamental inconsistencies. First, it has to be reconciled with the phenomenological Landau theory of phase transitions, according to which the order parameter must be associated with a particular irreducible representation of the relevant symmetry group. Following this prescription, Volovik and Gor'kov developed a classification of the exotic superconducting phases within the GL theory\cite{Volovik} (a systematic classification of the GL theories based on the symmetry analysis can be found in Ref.~\onlinecite{ueda}). It is important that the number of independent order parameters in the GL theory, given by the dimensionality of the irreducible representation, is typically lower then the number of bands, which is certainly  different from the multicomponent model mentioned above.

Second, the analysis of the multi-component GL model presented by Geilikman, Zaitsev and Kresin\cite{geilik,kresin} and more recently by Kogan and Schmalian,\cite{kogan} revealed another inconsistency: the accuracy of a solution to the formalism {\it exceeds} the accuracy of its derivation. This discrepancy is intrinsic in the multiband generalization of the Gor'kov procedure and can only be eliminated by invoking an additional truncating {\it reconstruction}, which removes the artificial higher-order contributions.\cite{geilik,kresin,kogan,extGL1,extGL2}

Without additional symmetries such reconstruction yields a {\it strict proportionality} of all band gaps, i.e., the GL theory has a single order parameter. This conclusion agrees with the phenomenological classification that predicts a single-component GL theory in this case. (Deviations from this result appear only in higher-order corrections to the ordinary GL theory.\cite{extGL1,extGL2,vagov})

This analysis did not consider the case of a degenerate solution for $T_c$ which appears due to an additional symmetry of the system. Furthermore, the calculations in Ref.~\onlinecite{geilik} for an arbitrary number of bands employed a rather restrictive ansatz for the band gaps, while Refs.~\onlinecite{kogan,extGL1} utilized the separability specific to the two-band case. The microscopic derivation of the multiband GL theory has not been yet achieved in the general case. Notice, that a mechanical merge of the symmetry analysis with the Gor'kov truncating procedure, in which the outcome of the Gor'kov procedure is simply rewritten in terms of the basis states of the relevant symmetry group representations, does not solve the problem. It yields a mixture of different irreducible representations, which should not happen in the standard GL formalism.\cite{gork1}

In this work we derive the reconstructed (true) GL theory from the microscopic Hamiltonian for a multiband superconductor in a general case with an arbitrary number of bands as well as with an arbitrary symmetry (reflected in the degeneracy of the solution for $T_c$). The origin of the symmetry is not important here. We note that it can appear not only due to the lattice structure of the material, as discussed in Ref.~\onlinecite{Volovik,ueda}, but also due to other reasons, e.g., the geometrical shape of the sample like in superconducting single-crystalline nanofilms.~\cite{nanofilms_int} A detailed analysis of the obtained equations is then performed for the three-band system treated as a prototype of a multiband superconductor. In particular, we consider a simple three-band model of pnictides with dominant interband couplings which allows for the two-fold degeneracy of the solution for $T_c$. We demonstrate that in full agreement with the phenomenological GL theory this system has two order parameters, related to the two-dimensional irreducible representation of the relevant symmetry group. However, unlike the phenomenological analysis based on the symmetry consideration, the derivation from the microscopic theory offers the explicit expressions for the coefficients of the GL theory. These expressions are highly nontrivial in the case of multiband superconductors because they contain important information about contributions of different bands that cannot be obtained from the symmetry arguments.~\cite{note} The corresponding ground state of the system is found to be a chiral state with a nontrivial phase difference between the band gaps. Such states in multigap superconductors have attracted much interest\cite{tes,frust,carl,tanaka} as they could lead to unconventional phenomena such as the formation of antiferromagnetic domains or noninteger vortices, see, e.g., Ref.~\onlinecite{ueda}. Notice that the present work does not go beyond the standard GL domain (i.e., band gaps are proportional to $\tau^{1/2}$, with $\tau=1-T/T_c$ the proximity to the critical temperature). An extended version of the multiband GL theory with the proper higher-order contributions to the band gaps will be published elsewhere.

The paper is organized as follows. In Sec.~\ref{sec:Gor} the GL theory for multiband superconductors is derived starting from the standard multiband BCS model. The derivation is performed in three steps: (i) the truncated multiband gap equation is obtained in matrix form by following the Gor'kov procedure adapted for the case of multiple bands, (ii) the truncating reconstruction is then applied by invoking the $\tau$-expansion, and (iii) an explicit form of the resulting GL equations is obtained by keeping the terms of order $\tau^{1/2}$ in the band gaps. In Sec.~\ref{sec:expl} we recast the final formalism in a more explicit form, for both the nondegenerate and degenerate cases. In Sec.~\ref{sec:N3} we consider a three-band model, for which expressions for the coefficients of the GL equations can be calculated analytically for an arbitrary interaction matrix. Then, we investigate the case of a degenerate solution for $T_c$ for a simple variant of the model with strong interband couplings and demonstrate analytically that the degeneracy in this model leads to the chiral ground state. Our summary and conclusions can be found in Sec.~\ref{sec:summ}.

\section{Derivation of the GL theory}
\label{sec:Gor}

\subsection{Truncated gap equation}
\label{subsec:trunc}

Following Gor'kov,\cite{gork} the GL theory is usually derived from the gap self-consistency equation, using an expansion of the anomalous Green's function in powers of the order parameter and its spatial derivatives. We outline this derivation for multiple bands, starting from the multiband BCS Hamiltonian\cite{suhl,mos} with the s-wave singlet pairing, which reads as
\begin{align}
H_{\rm BCS}=&H_c+\sum_i\int\!\!d^3 r\, \Bigl[ \sum_\sigma \hat\psi^{\dagger}_{i\sigma}({\bf r})\, T_i({\bf r})\hat\psi_{i\sigma}({\bf r})\notag\\ &+ \hat\psi^{\dagger}_{i\uparrow}({\bf r}) \,\hat\psi^{\dagger}_{i\downarrow}({\bf r})\,\Delta_i({\bf r})+\, {\rm h.c.} \Bigr], \label{eq:BCS}
\end{align}
where $i$ enumerates different bands, $\hat\psi_{i\sigma}({\bf r})$ and $\hat\psi^{\dagger}_{i\sigma}({\bf r})$ are the carrier field operators, $\Delta_i({\bf r})$ are the band gap functions, or simply band gaps, $H_c$ is the $c$-number term (see, e.g., Ref.~\onlinecite{zhit}), and $T_i({\bf r})$ stands for the single-electron energy. Equation~(\ref{eq:BCS}) is accompanied by the self-consistency gap equation
\begin{align}
\vec{\Delta}=\check{g}\vec{R},
\label{eq:gap}
\end{align}
where we introduce $\check{g}$, the matrix of the coupling constants $g_{ij}$, and use the vector notations $\vec{\Delta}^T= (\Delta_1,\Delta_2,\ldots)$ and $\vec{R}^T=(R_1,R_2,\ldots)$, with $R_i=\langle\hat\psi_{i\uparrow}({\bf r})\hat\psi_{i\downarrow}({\bf r})\rangle$ being the anomalous Green's function of the $i$-th band.~\cite{fett}

Using Eq.~(\ref{eq:BCS}), we expand the anomalous Green's functions in the vicinity of $T_c$ into a series in powers of the band gaps and their spatial gradients. As the Hamiltonian (\ref{eq:BCS}) is diagonal over the band index, the series is obtained independently for each band yielding the same expressions as in the single-band case. Referring interested readers to the original calculations,\cite{gork} here we quote the well-known final expansion for $R_i$, where only the leading nonlinear and gradient terms are retained,
\begin{align}
R_i[\Delta_i]\simeq N_i(0)\mathcal{A}\Delta_i +\Omega_i[\Delta_i],
\label{eq:R}
\end{align}
where $N_i(0)$ is the band DOS, ${\mathcal A} = {\rm ln} \big(\frac{2e^{\Gamma} \hbar\omega_c}{{\pi}T_{c}}\big)$, and
\begin{equation}
\Omega_i[\Delta_i]=-a_i\Delta_i - b_i|\Delta_i|^2\Delta_i + \mathcal{K}_i {\boldsymbol\nabla}^2\Delta_i.
\label{eq:Omeg}
\end{equation}
In Eqs.~(\ref{eq:R}) and (\ref{eq:Omeg}) $\omega_c$ is the cut-off frequency, $\Gamma=0.577$ is the Euler constant and the coefficients are calculated as
\begin{equation}
a_i= - N_i(0)\tau,\;b_i=N_i(0)\frac{7\zeta(3)}{8\pi^{2}T^2_c},\; \mathcal{K}_i=\frac{b_i}{6}\hbar^2v^2_i,
\label{eq:coeff}
\end{equation}
with $\zeta(\dots)$ the Riemann zeta function and $v_i$ the band Fermi velocity.
Although here only results for the clean limit are quoted, we note that the structure of the equations will be the same for dirty systems, as is usually the case in the standard GL formalism.\cite{disorder}

We note that the magnetic field is not included in Eq.~(\ref{eq:Omeg}). The generalization to the nonzero-field case is trivial and will be done on a later stage by using the standard prescription of inserting the gauge invariant gradient ${\bf D}=\boldsymbol\nabla-i\frac{2e}{\hbar c}{\bf A}$, where ${\bf A}$ is the vector potential. One should remember, however, that this recipe is valid exclusively for the standard GL domain when only terms of order $\tau^{1/2}$ are kept in the gap functions. A more involved and complex procedure is needed when higher-order corrections to the band gaps are incorporated.\cite{vagov}

Substituting Eq.~(\ref{eq:R}) into Eq.~(\ref{eq:gap}), we obtain the following system of coupled equations:
\begin{align}
\big(\mathcal{A}_i+ a_i\big)\Delta_i+b_i\Delta_i|\Delta_i|^2-\mathcal{K}_i {\boldsymbol\nabla}^{2}\Delta_i+\sum_{j\neq i}\gamma_{ij}\Delta_j=0, \label{multibandeq}
\end{align}
where $\gamma_{ij}$ is the element of the inverted interaction matrix $\check{g}^{-1}$ and the constants $\mathcal{A}_i$ are defined as
\begin{align}
\mathcal{A}_i=\gamma_{ii} - N_i(0)\mathcal{A}.
\label{coef}
\end{align}

The truncated equations given by Eg.~(\ref{multibandeq}) are commonly referred to as a generalization of the GL theory to the multiband case, or the {\em multi-component GL theory}. This interpretation is suggestive, especially given that in the limit of zero interband couplings Eq.~(\ref{multibandeq}) yields $N$ uncoupled GL equations for $\Delta_i$'s that are the true Landau order parameters for the uncoupled bands. Thus, the coupling is commonly assumed to be a weak perturbation that does not significantly alter this physical picture.

However, already this trivial limit highlights shortcomings of the interpretation of Eq.~(\ref{multibandeq}) as a consistent multiband GL formalism. In the absence of coupling each band has its own critical temperature $T_{ci}$ while $T_c$ of the entire system is the largest of those. In the vicinity of $T_c$, which is the usual validity domain of the GL theory, only the band with $T_{ci} = T_c$ develops a superconductive state and, therefore, the system is in fact described by a single-band GL theory with the single order parameter (here we assume that $T_{ci}$'s are well separated). One can also imagine a degenerate situation when $M\le N$ gaps have the same largest critical temperature $T_c$. Here, in the zero-coupling limit, the system is described by the theory with $M\le N$ order parameters corresponding to $M$ uncoupled components. Thus, in the zero interband-coupling limit, the GL theory {\em always} has fewer active order parameters then the number of the available bands. This conclusion is of course trivial in the noninteracting case. However, in what follows we demonstrate that it holds also in the general case of a nonzero coupling within the accuracy of the GL approach.

\subsection{Reconstructed GL theory}
\label{subsec:rec}

Deriving the GL theory for the general case of nonzero interband interactions starts by noting that as discussed in the introduction, Eq.~(\ref{multibandeq}) is inconsistent because the accuracy of its solution exceeds the accuracy of its derivation. One can see this (details can be found in Refs.~\onlinecite{geilik,kresin,kogan,extGL1,extGL2}) by taking into account that the coefficients $\mathcal{A}_i+a_i$ and $\gamma_{ij}$ are not zero in the limit $\tau \to 0$ ($T\to T_c$). This implies that a solution to Eq.~(\ref{multibandeq}), when being expanded in $\tau$, comprises terms of arbitrarily high orders, i.e., all $\Delta_i$'s are given by infinite series in powers $\tau^{n+1/2}$, with integer $n$. At the same time the Gor'kov truncation neglects terms that contribute to orders higher than $\tau^{1/2}$ in the band gap functions. The only situation when this inconsistency does not happen is the single-band GL theory where the coefficient of the linear term in the GL equation is proportional to $\tau$ and, as a result, the solution comprises a single contribution of order $\tau^{1/2}$.

In order to reconcile the accuracy of a solution for $\Delta_i$ with the accuracy of the derivation of Eq.~(\ref{multibandeq}), we use the reconstruction procedure that abandons incomplete higher-order contributions from the band gaps. This procedure is nothing more than a systematic perturbation expansion in $\tau$, which gives the GL theory and its corrections in a systematic way.\cite{vagov} Following this procedure, the solution to Eq.~(\ref{multibandeq}) is sought in the form of a series in odd powers of $\tau^{1/2}$ as
\begin{align}
\Delta_i=\Delta^{(0)}_i + \Delta^{(1)}_i +\mathcal{O}(\tau^{5/2}),
\label{expansion}
\end{align}
where $\Delta^{(0)}_i \propto \tau^{1/2}$ and $\Delta^{(1)}_i \propto \tau^{3/2}$.
This series is inserted into Eq.~(\ref{multibandeq}) and then the terms of the same order are collected. A simple power-counting shows that making a solution to Eq.~(\ref{multibandeq}) consistent with the Gor'kov truncation, one should keep the two lowest orders in the resulting $\tau$-expansion of Eq.~(\ref{multibandeq}) and the leading order term $\propto \tau^{1/2}$ in Eq.~(\ref{expansion}).

Notice that one must also take into account that spatial derivatives of the difference $\delta \Delta_i = \Delta_i-\Delta^{(0)}_i$ do not contribute to these lowest orders. In an earlier consideration~\cite{geilik} it was assumed that $\delta \Delta_i$ is independent of the coordinates, see Eq.~(14) in Ref.~\onlinecite{kresin}. Subsequent works\cite{extGL1,extGL2,vagov} have demonstrated that such a restrictive ansatz is not needed. The GL theory introduces the coherence length, $\xi \propto \tau^{-1/2}$ so that {\em all} spatial derivatives of {\em all} contributing terms in the band gaps scale as $\propto \tau^{1/2}$. In other words, each gradient operator $\nabla$ introduces a factor $\propto \tau^{1/2}$. Counting powers of $\tau$ in the expansion confirms that the higher-order gradients of $\Delta_i^{(0)}$ as well as the lowest gradients of $\delta\Delta_i$ do not contribute into the two lowest orders of the $\tau$-expansion of Eq.~(\ref{multibandeq}).

Substituting Eq.~(\ref{expansion}) into Eq.~(\ref{multibandeq}) and collecting the terms of order $\tau^{1/2}$ we obtain the first equation in the reconstructed theory
\begin{equation}
\check{L}\vec{\Delta}^{(0)}=0,
\label{eq:1/2}
\end{equation}
where elements of matrix $\check{L}$ are written as
\begin{align}
L_{ij} = \delta_{ij} {\cal A}_i + (1-\delta_{ij})\gamma_{ij}
\label{eq:matrixL}
\end{align}
with $\delta_{ij}$ being the Kronecker symbol. The condition of solvability of Eq.~(\ref{eq:1/2}),
\begin{align}
{\rm det}\check{L}=0,
\label{eq:Tc}
\end{align}
is the equation for the critical temperature $T_c$ that generally has $N$ solutions. Clearly, one has to choose the solution with the maximal $T_c$ as it yields the minimal value of the free energy. Equation~(\ref{eq:1/2}) is commonly referred to as the linearized gap equation as it can also be obtained by simply neglecting all the nonlinear contributions in Eq.~(\ref{multibandeq}).

When $N > 2$, one may encounter a situation with $M < N$ degenerate solutions to Eq.~(\ref{eq:Tc}) that correspond to the same maximal value of $T_c$. In this case the matrix $\check L$ has $M$ eigenvectors $\xi_\alpha$, with $\alpha = 1,...,M$, corresponding to the zero eigenvalue of $\check L$ at $T=T_c$. Without loss of generality these eigenvectors can be chosen orthogonal, and their normalization is not important.

A general solution to Eq.~(\ref{eq:1/2}) is then represented as a sum of $M$ terms ($M=1$ is for the nondegenerate case), one for each eigenvector, as
\begin{align}
\vec{\Delta}^{(0)}({\bf r})=\sum\limits_\alpha \psi_{\alpha}({\bf r}) \vec\xi_{\alpha}.
\label{eq:1/2deg}
\end{align}
Here $M$ functions $\psi_{\alpha}$ are specified by the equation that is obtained from Eq.~(\ref{multibandeq}) by matching terms of order $\tau^{3/2}$ as
\begin{align}
\check{L}\vec{\Delta}^{(1)} = \vec{\Omega} [\vec\Delta^{(0)}],
\label{eq:3/2}
\end{align}
where the components $\Omega_i[\Delta_i^{(0)}]$ of $\vec{\Omega}[\vec \Delta^{(0)}]$ are given by Eq.~(\ref{eq:Omeg}) with $\Delta_i$ replaced by $\Delta_i^{(0)}$. A closed set of $M$ equations for $\psi_\alpha({\bf r})$ is derived by projecting  Eq.~(\ref{eq:3/2}) to the eigenvectors $\vec\xi_{\alpha}$, which yields $M$ equations given by
\begin{align}
\sum\limits_i\xi_{\alpha i}\Omega_i[\Delta_i^{(0)}]=0,
\label{eq:3/2deg}
\end{align}
where $\xi_{\alpha i}$ is the $i$-th component of $\vec{\xi}_\alpha$.

\section{Explicit form of the GL equations}
\label{sec:expl}

\subsection{Nondegenerate case}

Here we recast Eq.~(\ref{eq:3/2deg}) in a more explicit and familiar form. In
the nondegenerate case a single function $\psi({\bf r}) \equiv
\psi_1({\bf r}) $ controls the same spatial profile of all band
condensates. Rewriting Eq.~(\ref{eq:3/2deg}) for $\psi({\bf r})$ one
obtains
\begin{equation}
a\psi+b|\psi|^2\psi-\mathcal{K}{\bf D}^{2}\psi=0, \label{eq:3/2nondegA}
\end{equation}
where we include a nonzero magnetic field by replacing ${\boldsymbol\nabla} \to {\bf D}$. The coefficients $a,\,b$ and $\mathcal{K}$ in Eq.~(\ref{eq:3/2nondegA}) are given by
\begin{equation}
a=\sum\limits_i a_i \xi_i^2,\,b=\sum\limits_i b_i \xi_i^{4}, \,
\mathcal{K}=\sum\limits_i\mathcal{K}_i\xi_i^2,
\label{eq:coef1}
\end{equation}
where $\xi_i$ is the band component of $\vec\xi \equiv \vec\xi_1$. The corresponding free-energy functional reads as
\begin{align}
\mathcal{F}=\int\!d^3r\,\Big[a|\psi|^2+\frac{b}{2}|\psi|^4+\mathcal{K}|{\bf D}\psi|^2+\frac{{\bf B}^2}{8\pi} \Big].
\label{eq:functnondeg}
\end{align}
Using this functional one derives the accompanying Maxwell equation for the gauge field in the form
\begin{equation}
\frac{1}{4\pi}{\rm rot}{\bf B}=i\frac{2e}{\hbar c} \mathcal{K}\big(\psi{\bf D}^\ast\psi^\ast -  \psi^\ast{\bf D}\psi\big).
\label{eq:Maxnondeg}
\end{equation}
As seen, Eqs.~(\ref{eq:3/2nondegA}), (\ref{eq:functnondeg}) and (\ref{eq:Maxnondeg}) have the form of the ordinary single-band GL theory. In fact, however, this is an {\it effectively} single-band GL theory as the coefficients $a,\,b$ and $\mathcal{K}$ comprise contributions of all bands. It is also important to remember that $\psi$ itself cannot be interpreted as an excitation gap: it is related to the band gap functions via Eq.~(\ref{eq:1/2deg}).

The single-band representation of the reconstructed GL theory allows one to define the characteristic lengths of a multiband superconductor in a unique way. In particular, the coherence length $\xi$, the magnetic penetration depth $\lambda$ and their ratio $\kappa$ are given by the standard GL expressions as
\begin{equation}
\xi=\sqrt{\frac{\mathcal{K}}{|a|}}, \quad
\lambda = \frac{\hbar c}{|e|} \sqrt{\frac{b}{32\pi\mathcal{K}|a|}}, ~ \kappa=\Phi_0\sqrt{\frac{b}{32\pi^3\mathcal{K}^2}}
\label{eq:xilnondeg}
\end{equation}
where $\Phi_0$ is the flux quantum. However, the multiband origin of Eqs.~(\ref{eq:3/2nondegA})-(\ref{eq:Maxnondeg}) is still reflected in some properties of the system. For example, following Eqs.~(\ref{eq:coef1}) and (\ref{eq:xilnondeg}), one concludes that $b,\mathcal{K}$ can be roughly estimated as linearly proportional to $N$. Taking into account the relation $\kappa \propto \sqrt{b/\mathcal{K}^2}$ one arrives at the trend $\kappa \sim 1/\sqrt{N}$, which means that a multiband superconductor should approach the type-I character when the number of bands is large enough.

\subsection{Degenerate case}

When the maximal solution to Eq.~(\ref{eq:Tc}) is degenerate, i.e.,
$M>1$, Eq.~(\ref{eq:1/2deg}) yields a set of coupled nonlinear
equations, an explicit form of which is obtained as
\begin{align}
\sum\limits_{\beta} \big(a_{\alpha\beta} -\mathcal{K}_{\alpha \beta} {\bf D}^2\big)\psi_\beta +\sum\limits_{\beta\gamma\delta} b_{\alpha \beta \gamma \delta}\,\psi_{\beta}\psi_{\gamma}^\ast\psi_{\delta}=0,
\label{eq:3/2degA}
\end{align}
where the coefficients are defined as
\begin{align}
&a_{\alpha \beta} =\sum\limits_i a_i \xi_{\alpha i} \xi_{\beta i},  \quad   \mathcal{K}_{\alpha \beta} =\sum\limits_i \mathcal{K}_i \xi_{\alpha i} \xi_{\beta i} ,\notag \\
& b_{\alpha\beta\gamma\delta}=\sum\limits_i b_i \xi_{\alpha i} \xi_{\beta i} \xi_{\gamma i} \xi_{\delta i}.
\label{eq:coef2}
\end{align}

The corresponding free-energy functional is now obtained in the form
\begin{align}
\mathcal{F} =&\int\!d^3r\,\Big[\sum\limits_{\alpha\beta} \big( a_{\alpha\beta} \psi^\ast_\alpha \psi_\beta +\mathcal{K}_{\alpha \beta}
{\bf D}^\ast\psi_\alpha^\ast\,{\bf D}\psi_\beta\big)\notag\\
&+\frac{1}{2}\sum\limits_{\alpha\beta\gamma\delta}
b_{\alpha \beta\gamma\delta}\; \psi_\alpha^\ast\,\psi_{\beta}\,\psi_{\gamma}^\ast\, \psi_{\delta} + \frac{{\bf B}^{2}}{8\pi} \Big].
\label{eq:functdeg}
\end{align}
By calculating the functional derivative with respect to the vector potential, we obtain from Eq.~(\ref{eq:functdeg}) the accompanying Maxwell equation as
\begin{equation}
\frac{1}{4\pi}{\rm rot}{\bf B}=i\frac{2e}{\hbar c} \sum\limits_{\alpha\beta}\mathcal{K}_{\alpha\beta}\big(\psi_\alpha{\bf D}^\ast\psi^\ast_{\beta} -  \psi^\ast_{\beta}{\bf D}\psi_\alpha\big).
\label{eq:Maxdeg}
\end{equation}

The number of components in the reconstructed GL theory is $1\le M < N$, unlike in the original system of equations given by Eq.~(\ref{multibandeq}). Another important difference is that all coefficients of the linear terms in the reconstructed GL theory are now proportional to $\tau$, which dictates that $\psi_{\alpha} \propto \tau^{1/2}$. This eliminates the problem of the mismatch between the accuracy of the solutions and equations, which was the reason to seek the reconstruction.

Notice that the reconstructed GL formalism, obtained here by the $\tau$-expansion, recovers the standard Landau theory of phase transitions. In particular, the degeneracy of the linearized gap equation is related to an extra symmetry between bands, hidden in the relevant coupling matrix. The degree of degeneracy $M$ is defined by the dimensionality of the corresponding irreducible representation with the $M$ basis vectors $\xi_{\alpha}$'s. Equation (\ref{eq:functdeg}) is interpreted as the Landau free-energy functional with $\psi_\alpha$'s being Landau order parameters. The reconstruction can thus be regarded as the procedure of finding the true Landau order parameter of the system, in the form of linear combinations of the band gaps, see Eq.~(\ref{eq:1/2deg}). However, Eqs.~(\ref{eq:functnondeg}) and (\ref{eq:functdeg}) are derived by matching all relevant terms in the $\tau$-expansion, rather than through the phenomenological approach based on the group-theory analysis.\cite{ueda}

In agreement with the Landau recipe, the reconstructed GL theory is based on a single irreducible representation. However, if one continues the $\tau$-expansion to next orders, admixtures of other irreducible representations will appear in the formalism. Within the symmetry analysis, it is often argued that such terms should arise because the appearance of the condensate at $T < T_c$ already changes the symmetry of the system.\cite{ueda} The reconstruction yields a clear quantitative estimate for such admixtures. It is easy to see from Eq.~(\ref{expansion}) that the order parameters related to extra irreducible representations will be of order $\tau^{3/2}$ and higher, which must be neglected in the present analysis concerning the standard GL formalism.

\section{Three band system}
\label{sec:N3}

\subsection{Eigenvectors}
\label{sec:N3a}

As a prototype of multiband superconductors, we now consider a physically relevant case of a three-band system, the analysis of which can be done in the analytical form. In order to obtain the eigenvectors $\xi_\alpha$, we write Eq.~(\ref{eq:1/2}) as a system of linear algebraic equations
\begin{eqnarray}
\mathcal{A}_i\Delta^{(0)}_i+\sum_{j\neq i}\gamma_{ij}\Delta^{(0)}_j=0.
\label{eq:1/2A}
\end{eqnarray}
It is easy to verify that the following relations hold
\begin{eqnarray}
\eta_1\Delta^{(0)}_1=\eta_2\Delta^{(0)}_2=\eta_3\Delta^{(0)}_3, \label{eq:1/2B}
\end{eqnarray}
where
\begin{align}
&\eta_1 = \mathcal{A}_1\gamma_{23}-\gamma_{12}\gamma_{13},\;
\eta_2 = \mathcal{A}_2\gamma_{13}-\gamma_{12}\gamma_{23},\notag\\
&\eta_3 = \mathcal{A}_3\gamma_{12}-\gamma_{13}\gamma_{23},
\label{eq:delta}
\end{align}
and $\mathcal{A}_i$ is defined by Eq.~(\ref{coef}).

We now investigate the following possibilities. Let us first assume that $\eta_1,\eta_2,\eta_3\not=0$. Then, from Eq.~(\ref{eq:1/2B}) we immediately find that
\begin{equation}
\xi_i \propto 1/\eta_i,
\label{eq:xicomp}
\end{equation}
which implies that the gaps in all three bands are nonzero. When one of the $\eta_i$'s is zero, say $\eta_1=0$, then Eq.~(\ref{eq:1/2B}) dictates that $\Delta^{(0)}_2= \Delta^{(0)}_3 = 0$, and therefore the condensate is formed only in one band. When two of the $\eta_i$'s vanish, the gap is nonzero in the corresponding two bands. In all these cases we deal with the nondegenerate scenario governed by the single-component GL equation (\ref{eq:3/2nondegA}) with the coefficients given by Eq.~(\ref{eq:coef1}). However, the eigenvector $\vec{\xi}$, whose band components appear in Eq.~(\ref{eq:coef1}), is dependent on a particular situation. As mentioned above, for $\eta_1,\eta_2,\eta_3\not=0$ we obtain Eq.~(\ref{eq:xicomp}) whereas for, say, $\eta_1=\eta_2=0$ we have $\vec{\xi}^{\,T}=(1,-\gamma_{13}/\gamma_{23},0)$.

The case when all $\eta_i$'s are equal to zero requires a bit more algebra. Expressing $\mathcal{A}_i$ in terms of $\gamma_{ij}$ from Eq.~(\ref{eq:delta}) and then inserting the result into Eq.~(\ref{eq:1/2A}), we find that in this case Eqs. (\ref{eq:1/2A}) are reduced to a single equation that reads as
\begin{align}
\gamma_{12}\gamma_{13}\Delta^{(0)}_1 + \gamma_{12}\gamma_{23}\Delta^{(0)}_2 +
\gamma_{13}\gamma_{23}\Delta^{(0)}_3=0.
\label{eq:1/2degA}
\end{align}
A general solution to Eq.~(\ref{eq:1/2degA}) can be written as
\begin{align}
\vec{\Delta}^{(0)}({\bf r})= \vartheta_1({\bf r})\vec{u}_1 + \vartheta_2({\bf r})\vec{u}_2,
\label{eq:1/2degC}
\end{align}
where
\begin{align}
\vec{u}_1=
\left(\begin{array}{c}
0 \\ -\gamma_{13}/\gamma_{12} \\ 1
\end{array}\right),
\;\vec{u}_2 =
\left( \begin{array}{c}
1 \\ -\gamma_{13}/\gamma_{23} \\ 0
\end{array}
\right)
\end{align}
are linearly independent and $\vartheta_{1,2}({\bf r})$ are unknown functions to be specified later. Comparing Eq.~(\ref{eq:1/2degC}) with Eq.~(\ref{eq:1/2deg}), we conclude that this case represents the degenerate scenario with $M=2$. Equation~(\ref{eq:1/2degC}) can be rewritten in terms of two orthogonal eigenvectors $\vec{\xi}_{1,2}$ by applying the orthogonalization procedure to $\vec{u}_{1,2}$, which gives
\begin{align}
\vec{\xi}_1=\vec{u}_1,\;\vec{\xi}_2 = \vec{u}_2 - \frac{\gamma^2_{13} \gamma_{12}}{(\gamma^2_{12} +\gamma^2_{13})\gamma_{23}}\,
\vec{u}_1.
\label{eq:1/2degD}
\end{align}
One can then express the functions $\vartheta_{1,2}$ through $\psi_{1,2}$ introduced earlier as $\vartheta_1({\bf r})=\psi_1({\bf r}) - \gamma^2_{13} \gamma_{12}/[(\gamma^2_{12} + \gamma^2_{13})\gamma_{23}]\psi_2({\bf r})$ and $\vartheta_2({\bf r})=\psi_2({\bf r})$. The band gaps are then defined by the two Landau order parameters $\psi_{1,2}$ according to
\begin{align}
&\Delta^{(0)}_1=\psi_2, \notag \\
&\Delta^{(0)}_2=-\frac{\gamma_{13}}{\gamma_{12}} \psi_1-\frac{\gamma^2_{12}\gamma_{13}}{(\gamma^2_{13} +\gamma^2_{12})\gamma_{23}}\psi_2, \notag \\
&\Delta^{(0)}_3=\psi_1-\frac{\gamma^2_{13}\gamma_{12}}{(\gamma^2_{13} +\gamma^2_{12})\gamma_{23}}\psi_2.
\label{eq:Delta_3_band}
\end{align}

Finally we note that the derivation of Eqs.~(\ref{eq:1/2degA})-(\ref{eq:Delta_3_band}) assumes that $\gamma_{12},\gamma_{13}, \gamma_{23}\not=0$. If some of these interband couplings is zero while $\eta_1 =\eta_2 =\eta_3=0$, the problem reduces to a trivial example of the nondegenerate case where some of the available bands are uncoupled.

\subsection{Chiral state with phase frustration}
\label{sec:N3b}

Under certain conditions the ground state of a three-band superconductor may develop a nontrivial phase difference between different band gaps, referred to as the state with the phase frustration or the chiral solution. This state is of a particular
interest as it breaks the time-reversal invariance in the system, leading to many unconventional superconducting properties.\cite{ueda} Below we analytically demonstrate the possibility of such a state in the three-band system within the simple variant of the model with strong interband couplings, i.e., $g_{ii}=0$ and $g_{i\neq j} >0$ and $N_1(0)=N_2(0)=N_3(0)$. Our analytical consideration compliments numerical investigations in the recent Ref.~\onlinecite{tes}. Such a model describes an interesting example of a system where the superconducting pairing is caused by the interband coupling and, as it is believed, may be relevant for pnictides.\cite{tes} We are interested in the special case when different interband couplings are equal to one another, which may be dictated by some symmetry between bands\cite{gork1} but is not necessarily limited to only this physical situation. Please note that many different combinations of intra- and interband couplings can lead to a degeneracy of $T_c$ and possible phase frustration (see, e.g. Ref.~\onlinecite{tanaka}). However, in the absence of physical justifications for such coupling matrices, we refrain from their analysis.

Using the orthogonality conditions for $\vec{\xi}_\alpha$'s and the fact that the band DOS's are equal, we obtain $a_{12}=a_{21}=0$ in Eq.~(\ref{eq:3/2degA}).  Furthermore, it is obvious that the tensor $b_{\alpha\beta\gamma\delta}$ is symmetric with respect to the permutation of each pair of the indices so that it is convenient to introduce new notations
\begin{align}
&\beta_1 =  b_{1 1 1 1}, \quad \beta_2 =  b_{1 1 1 2} =b_{1 1 2 1}= b_{1 2 1 1}= b_{2 1 1 1}, \notag \\
&\beta_3 =  b_{1 1 2 2} =b_{1 2 1 2} = b_{2 1 1 2}= b_{2 1 2 1}= b_{1 2 2 1}= b_{2 2 1 1}, \notag \\
&\beta_4 =  b_{1 2 2 2} =b_{2 1 2 2} =  b_{2 2 1 2} =b_{2 2 2 1}, \quad \beta_5 =  b_{2 2 2 2}.
\end{align}
Then, for a homogeneous case without a magnetic field, Eq.~(\ref{eq:3/2degA}) yields
\begin{subequations}\label{eq:threebanddeg}
\begin{align}
\alpha_1 =& -\beta_1|\psi_1|^2-\beta_2\big(2\psi_1^\ast\psi_2 +\psi_1\psi_2^\ast\big)\notag\\
&-\beta_3\Big(2|\psi_2|^2 + \psi^2_2\frac{\psi_1^\ast}{\psi_1}\Big)-\beta_4 |\psi_2|^2\frac{\psi_2}{\psi_1},\label{eq:Athreebanddeg}\\
\alpha_2 =& - \beta_2|\psi_1|^2\frac{\psi_1}{\psi_2} - \beta_3\Big(2|\psi_1|^2+\psi_1^2 \frac{\psi_2^\ast}{\psi_2}\Big)\notag\\
&-\beta_4\big(2\psi_1\psi^\ast_2+\psi^\ast_1\psi_2\big)-\beta_5|\psi_2|^2,
\label{eq:Bthreebanddeg}
\end{align}
\end{subequations}
where we also denote $\alpha_1=a_{11}$, $\alpha_2=a_{22}$. As usual, it is convenient to search for a solution to Eq.~(\ref{eq:threebanddeg}) in the form $\psi_i = |\psi_i|\exp(i\phi_i)$. Then, matching the imaginary parts in Eq.~(\ref{eq:Athreebanddeg}) [or in Eq.~(\ref{eq:Bthreebanddeg}), which gives the same result], we obtain
\begin{subequations}\label{eq:imreal}
\begin{align}
\big[\beta_2 r + 2\beta_3\cos(\delta\phi) + \beta_4 r^{-1}\big]\sin(\delta\phi)=0,
\label{eq:im}
\end{align}
where the notations $\delta\phi =\phi_2-\phi_1$ and $r = |\psi_1|/|\psi_2|$ are introduced. Matching the real parts in Eqs.~(\ref{eq:Athreebanddeg}) and (\ref{eq:Bthreebanddeg}) yields, respectively,
\begin{align}
-\frac{\alpha_1}{|\psi_1|^2} =& \beta_1+\big(3\beta_2 +\beta_4 r^{-2}\big)r^{-1}\cos(\delta\phi)\notag\\
&+\beta_3 r^{-2}\big[2+\cos(2\delta\phi)\big], \label{eq:real1} \\
-\frac{\alpha_2}{|\psi_1|^2} =& \big(\beta_2 r + 3\beta_4 r^{-1}\big) \cos(\delta\phi)\notag\\
&+\beta_3\big[2+\cos(2\delta\phi)\big]+\beta_5 r^{-2}.
\label{eq:real2}
\end{align}
\end{subequations}

To check the thermodynamic stability of different solutions to Eqs.~(\ref{eq:im}) and (\ref{eq:imreal}), one needs to calculate the free energy from the functional in Eq.~(\ref{eq:functdeg}). It can be rewritten, using the new notations, as
\begin{align}
\mathcal{F} =&\int\!d^3r\,\Big\{|\psi_1|^2\big(\alpha_1 +\alpha_2\,r^{-2}\big)+\frac{1}{2}|\psi_1|^4\notag\\
&\times\Big(\beta_1+ 4\cos(\delta\phi) r^{-1}\big(\beta_2+\beta_4r^{-2}\big)\notag\\
&+2\beta_3r^{-2}
\big[2+\cos(2\delta\phi)\big] + \beta_5 r^{-4}\Big)\Big\}.
\label{eq:free_frA}
\end{align}

To proceed further, we substitute the chosen model parameters into the obtained equations. Inverting the coupling matrix yields $\gamma_{ii} = -1/(2g)$ and $\gamma_{i\neq j}=1/(2g)$. Then, using Eq.~(\ref{eq:1/2degD}), we obtain the eigenvectors as $\vec{\xi}^{\,T}_1=(0,-1,1)$ and $\vec{\xi}^{\,T}_2=(2,-1,-1)$, where $\vec{\xi}_2$ is now multiplied by $2$ for the sake of convenience of our further calculations. Substituting these eigenvectors into Eq.~(\ref{eq:coef2}), we find
\begin{align}
&\alpha_1=2\tilde{a},\;\alpha_2 = 6\tilde{a},\; \beta_1 = 2\tilde{b},\; \beta_2=0,
\notag\\
&\beta_3=2\tilde{b},\;\beta_4=0,\; \beta_5=18\tilde{b},
\label{eq:coef3}
\end{align}
where $\tilde{a}=a_1=a_2=a_3$ and $\tilde{b}=b_1=b_2=b_3$, and $a_i$
and $b_i$ are given by Eq.~(\ref{eq:coeff}). Finally, based on Eq.~(\ref{eq:coef3}), we can rewrite Eq.~(\ref{eq:im}) as
\begin{align}
\sin(2\delta\phi)=0,
\label{eq:imA}
\end{align}
which yields the obvious solution for the phase difference $\delta\phi=\pi n/2$, with $n$ being integer.

One can identify two solution classes. The first one is given by $\delta\phi=0,\pi, 2\pi,\ldots$ at which $\cos(2\delta\phi)=1$. Here a sign difference can occur between the band components but there is no nontrivial phase difference. In this case Eqs.~(\ref{eq:real1}) and (\ref{eq:real2}) are reduced to
\begin{align}
|\psi_1|^2= |\tilde{a}|\big/\big[\tilde{b}\,(1+3r^{-2})\big].
\label{eq:real_ord}
\end{align}
The complete homogeneous solution for the band gaps is then given by
\begin{align}
\vec{\Delta}^{(0)}=\sqrt{\frac{|\tilde{a}|}{\tilde{b}(3+r^2)}}
\left(
\begin{array}{c}
2\\
-r-1\\
r-1
\end{array}
\right),
\label{eq:Delta_ord}
\end{align}
where $r=|\psi_1|/|\psi_2|$ serves as a parameter. The corresponding free-energy density $f=\mathcal{F}/V$ is obtained as
\begin{align}
f=-\tilde{a}^2/\tilde{b}.
\label{eq:free_ord}
\end{align}
Notice that since Eq.~(\ref{eq:free_ord}) does not depend on $r$, this quantity is not fixed and therefore we obtain a continuous family of solutions with the same free-energy density.

The second solution class is obtained when $n$ is odd, i.e., $\delta\phi= \pi/2,3\pi/2,\ldots$ and $\cos(2\delta\phi)=-1$. In this case Eqs.~(\ref{eq:real1}) and (\ref{eq:real2}) yield the system of two equations
\begin{subequations}\label{eq:real_fr}
\begin{align}
-\tilde{a}/|\psi_1|^2=&\tilde{b}\,(1+r^{-2}),
\label{eq:real_fr_1}\\
-3\tilde{a}/|\psi_1|^2=&\tilde{b}\,(1+9r^{-2}).
\label{eq:real_fr_2}
\end{align}
\end{subequations}
This system is solved trivially giving $|\psi_1|^2=3\tilde{a}/(4\tilde{b})$ and $r=\sqrt{3}$. Then, using Eqs.~(\ref{eq:real_fr}) and taking $\delta\phi=\pi/2,5\pi/2,\ldots$ and $\delta\phi=3\pi/2,7\pi/2,\ldots$ we obtain two different solutions as
\begin{align}
\vec{\Delta}^{(0)}=i\sqrt{\frac{|\tilde{a}|}{\tilde{b}}}
\left(
\begin{array}{c}
1\\
e^{i2\pi/3}\\
e^{-i2\pi/3}
\end{array}
\right);~ - i\sqrt{\frac{|\tilde{a}|}{\tilde{b}}}
\left(
\begin{array}{c}
1\\
e^{-i2\pi/3}\\
e^{i2\pi/3}
\end{array}
\right).
\label{eq:Delta_fr1}
\end{align}
These are chiral solutions with a nontrivial phase difference between the band gaps. The free-energy density for both of them reads as
\begin{align}
f=-3\tilde{a}^2/(2\tilde{b}).
\label{eq:free_fr}
\end{align}
Comparing this with Eq.~(\ref{eq:free_ord}) reveals that the chiral solution is more favorable energetically and thus the three-band model with strong interband couplings supports the formation of the chiral state.

This conclusion agrees with numerical simulations of Eq.~(\ref{multibandeq}) for the three-band case,\cite{tes} which showed that the chiral state with the phase shifts $\pm 2\pi/3$ is found at $T\to T_c$ only in the limit $g_{23} \to g_{12}=g_{13}$.
The phase shift obtained in our work is independent of temperature, which differs from numerical simulations in Ref.~\onlinecite{tes}. We note, however, that these simulations employed the unreconstructed GL equations, where a solution does not account for all relevant terms of the $\tau$-expansion. A correct temperature dependence of the phase shift must be calculated with the help of the extended multiband GL formalism that should be constructed in the spirit of the approach in Ref.~\onlinecite{extGL2}.

As already mentioned above, the appearance of the chiral state may indicate the symmetry of the model, reflected in the structure of the coupling matrix. In particular, the matrix investigated in this section can be realized by choosing the bands as the pockets of the Fermi surface centered around {\bf X} points of the Brillouin zone of the $fcc$ lattice, see Ref.~\onlinecite{gork1}. The band gaps are then transformed according to a three dimensional representation of the {\em O$_{\rm h}$} cubic symmetry group. This representation splits into one dimensional {\em A$_{\rm g}$} and two-dimensional {\em E$_{\rm g}$} irreducible representations. The two-dimensional representation {\em E$_{\rm g}$}, that corresponds to the highest critical temperature, can have its basis chosen as two vectors in Eq.~(\ref{eq:Delta_fr1}). Constructing the Landau theory from this irreducible representation, one recovers the reconstructed GL formalism discussed above, which additionally proves its validity.

Here we stress that recasting the multicomponent GL theory (\ref{multibandeq}) in terms of the basis functions of the symmetry-group representations does not eliminate admixtures of different irreducible representations in the free-energy functional.\cite{gork1} However, following our analysis, such admixtures must be neglected as they exceed the accuracy of the GL theory, in full agreement with the standard Landau approach. The proper accounting of the admixture terms can be done only by employing the extended GL theory which collects all relevant higher-order terms in the expansion of the band gaps.

Finally, as the chiral state is related to the degeneracy of a solution for $T_c$ that can be caused by, e.g., the crystalline symmetry, the existence of any simple relation between the chiral state and signs of the interband couplings $\gamma_{i\neq j}$, as suggested in Refs.~\onlinecite{carl,tanaka}, appears to be very questionable at least in the GL domain. Notice that this conclusion is also supported by numerical investigations of Eqs.~(\ref{multibandeq}) performed in Ref.~\onlinecite{tes}.

\section{Summary and conclusions}
\label{sec:summ}

In this work we have derived the consistent GL theory from the multiband BCS Hamiltonian. The derivation applies a reconstruction procedure to the conventional Gor'kov truncation of the matrix gap equation. This reconstruction invokes the expansion in powers of $\tau$ and removes incomplete contributions to band gaps of orders higher than $\tau^{1/2}$, thus matching the accuracy of the gaps with that of the Gor'kov truncation.

When the solution for $T_c$ is not degenerate, we recover the earlier results of Refs.~\onlinecite{geilik,kogan,extGL1,extGL2} that the GL theory of a multiband superconductor maps onto a single-component GL formalism in which the spatial profiles of all band gaps are equivalent. However, this result is valid only in the standard GL domain, i.e., to the accuracy $\Delta_i\propto \tau^{1/2}$. Difference between the spatial profiles of the band gaps appears already in the leading correction to the GL theory.\cite{extGL1,extGL2}

If the solution for $T_c$ is degenerate, which appears due to a symmetry of the system, the GL theory acquires several order parameters. We have carried out a detailed analysis for the three-band system treated as a prototype of a multiband superconductor. For the simple three-band model of pnictides with dominant interband couplings, the solution for $T_c$ is two-fold degenerate and the GL theory has two order parameters $\psi_{\alpha}$ which correspond to the two-dimensional irreducible representation of the relevant symmetry group, in full agreement with the Landau theory. We have shown that the band energy gaps themselves cannot be interpreted as the Landau order parameters in a multiband superconductor due to the Josephson-like coupling between bands.

Our approach yields explicit expressions for the coefficients of the GL theory. Also, the formalism provides a solid basis for further extensions of the theory and, in particular, offers the correct way to account for the influence of other irreducible representations not inherent in the ordinary GL approach.

Although it was not a purpose of our work to discuss the origin of the degeneracy of $T_c$, it is worth noting that it does not always appear due to the crystalline symmetry. It can arise, e.g., in the atomically flat superconducting nanofilms, where the size quantization of the perpendicular motion of electrons results in the formation of multiple single-electron subbands.~\cite{nanofilms} Such superconducting nanofilms can be regarded as effectively-multiband superconductors with the interaction matrix~\cite{nanofilms_int} $g_{ij}=g(1+\delta_{ij}/2)/d$, where $d$ is the nanofilm thickness and $g$ is the coupling constant for the material of the nanofilm. The structure of this matrix is similar to that of $\check{g}$ investigated in Sec.~\ref{sec:N3b} and, as a result, a degenerate solution for $T_c$ also appears in this case. The developed formalism thus provides a general link between the multiband BCS theory and the phenomenological Landau model for multiband superconductors, irrespective of the origin of the symmetry.

We conclude by noting that the degenerate regime manifests itself in several important physical consequences such as the formation of the chiral ground state and the appearance of different spatial length-scales of the band condensates, which can be observed even at $T\to T_c$. This may result in a plethora of new phenomena, i.e., fractional vortices,\cite{babaev1} flux-carrying topological solitons,\cite{babaev2} and other exotic states.\cite{vortex_ex} So far those phenomena have been studied using the unreconstructed multi-component GL model given by Eqs.~(\ref{multibandeq}), and so we suggest revisiting these problems in the framework of the true GL formalism.

\begin{acknowledgments}
This work was supported by the ``Odysseus'' Program of the Flemish Government and the Flemish Science Foundation (FWO-Vl). A.A.S. acknowledges useful discussions with D. Neilson.
\end{acknowledgments}


\begin{thebibliography}{99}
\bibitem{disorder} We do not consider the case of an extremely strong disorder when additional nonlinear coupling terms can appear between the bands, see Ref.~\onlinecite{gur}.
\end{thebibliography}
\end{document}